\documentstyle[aps,epsfig,twocolumn,float]{revtex}

\newcommand{\vopo}{(VO)$_2$P$_2$O$_7$}
\restylefloat{figure}

\tighten
\begin{document}
\draft 
\title{A new model to describe the physics of \vopo }
\author{A.~Wei\ss e$^{1}$, G.~Bouzerar$^{2}$ and H.~Fehske$^{1}$}
\address{$^{1}$
Physikalisches Institut, Universit\"at Bayreuth, 95440 Bayreuth, Germany}
\address{$^{2}$
Institut f\"ur Theoretische Physik, Universit\"at zu K\"oln, 50937 K\"oln, Germany}
\address{~
  \parbox{14cm}{\rm 
    \medskip
    In the past different models for the magnetic salt vanadyl pyrophosphate
    \vopo{} were discussed. Neither a spin ladder nor an
    alternating chain are capable to describe recently measured magnetic
    excitations. In this paper we propose a 2D model that fits better to
    experimental observations.
}}

\maketitle

Low dimensional quantum spin systems have been a field of intense 
theoretical and experimental research over the last decades. Special interest
was given to spin ladder and chain materials. One compound that
has been examined in this context is the insulating magnetic salt
vanadyl pyrophosphate \vopo. Initially it was considered as a
prototypical realization of a two--leg antiferromagnetic Heisenberg
ladder~\cite{Johnston}. However, susceptibility data on polycrystalline and
single crystalline material could be well fitted with both, ladder or alternating
chain models~\cite{Johnston,Barnes,Schwenki}, stressing the fact that 
susceptibility is not too sensitive to the particular model. Early inelastic 
neutron scattering measurements on polycrystalline samples 
indicated a spin gap of about 3.7 meV and supported a two--leg 
ladder model with the coupling constants estimated from susceptiblity 
data~\cite{Eccleston}.

Recent neutron scattering experiments with powder samples~\cite{Garrett1} 
and with an array of single crystals~\cite{Garrett2} provided detailed 
information on the low--energy excitation spectrum. 
Garrett et al.~\cite{Garrett2} observed a triplet branch with strongest 
(antiferromagnetic) dispersion in b--direction, weak (ferromagnetic)
dispersion in a--direction, and a spin gap of 3.1~meV. 
Most notably they found an additional {\it second} branch, separated from 
the first by an energy smaller than the gap. 
This was inconsistent with the picture, of \vopo{} being a spin ladder in
a--direction, but also an alternating Heisenberg chain in b--direction
can not explain a second triplet branch over the whole Brillouin zone, 
as was shown recently~\cite{Bouzerar1,Barnes2}.

\begin{figure}[hbt]
\epsfig{file= 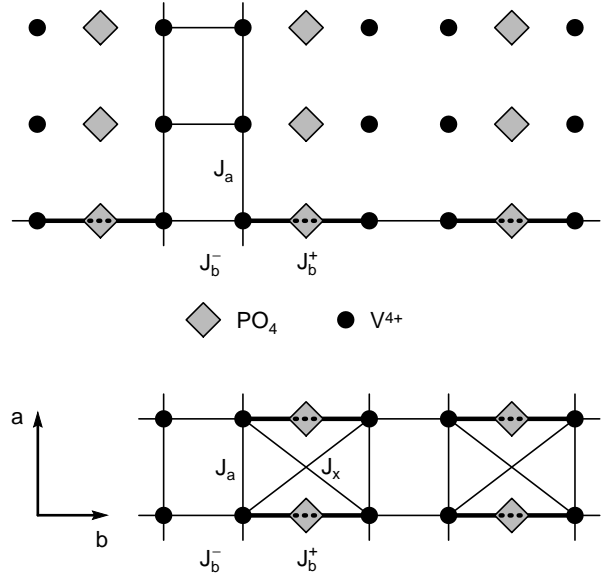, width = \linewidth}
\caption{Schematic structure of \vopo{}. The exchange couplings are 
depicted for 
(i) the ladder model $(J_{\parallel} = J_{a},J_{\perp}= J_{b}^{-})$, 
(ii) the alternating chain model $(J_{b}^{\pm}=J_b(1\pm\delta)$, 
and (iii) the new model $(J_b^{\pm},J_a,J_{\times})$.
Throughout we measure energies in units of $J_b$.}
\label{figstruct}
\end{figure}

In this work, starting with the alternating Heisenberg chain, we check
whether coupling of (two) chains resolves this puzzling situation. 
As we do not succeed proceeding this way, we consider a new, truly 
two--dimensional model. 
We perform exact diagonalizations of finite systems with up to 32 spins and
periodic boundary conditions, supplemented by finite--size analysis if 
possible.

The Hamiltonian of the alternating Heisenberg chain (AHC) reads as follows
\begin{equation}\label{ahc_ham}
H_{\rm AHC} = J_b \sum_i (1+\delta(-1)^i) {\bf S}_i\cdot{\bf S}_{i+1},
\end{equation}
where ${\bf S}_i$ are spin-$1\over 2$ operators and $i$ denotes the
sites in b--direction (see Fig.~\ref{figstruct}). For $\delta>0$ the 
spectrum has a gap; there is an one--magnon branch and a singlet branch, 
at least around momentum $\pi\over 2$, below a continuum of states.

\begin{figure}[tb]
\epsfig{file= 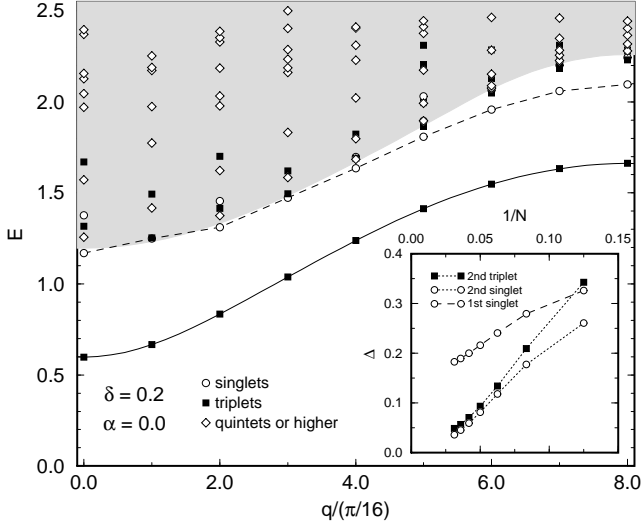, width = \linewidth, bbllx = 40, bblly = 40, bburx = 530, bbury = 450}
\caption{Low--energy excitations of the AHC. The inset shows the
difference $\Delta$ of the 1$^{\rm st}$ and 2$^{\rm nd}$ singlet and 
the 2$^{\rm nd}$ triplet to the 1$^{\rm st}$ quintet at momentum $\pi/2$ 
versus inverse chain length.}
\label{figahcdisp}
\end{figure}
As an example Fig.~\ref{figahcdisp} shows the low--lying excitations of 
a finite system of 32 sites for $\delta = 0.2$. The magnon branch 
is fitted to a sum of cosines $\omega_q^m = \sum_{n = 0}^5 a_n \cos(2 n q)$ and
the shaded region corresponds to the continuum of two--magnon excitations 
resulting from this dispersion.
Recently it was stressed~\cite{Bouzerar1,Uhrig} that there exists a second
well--defined triplet below the two--magnon continuum near momentum 
$\pi\over 2$, but as Fig.~\ref{figahcdisp} indicates, the second triplet 
occurs only very close to higher states, even for the relatively strong
dimerization of $\delta=0.2$. Therefore it was stated that an alternating 
chain will not explain the second triplet excitation observed in \vopo{} at 
{\it all} q--momenta. However, it is known~\cite{Bouzerar1} that including frustration, i.e. 
an antiferromagnetic  next nearest neighbor interaction $\alpha$ between
${\bf S}_i$ and ${\bf S}_{i+2}$, 
into the alternating chain model, yields a second well--defined triplet 
branch below the continuum in the whole Brillouin zone, provided 
$\alpha$ is sufficiently strong.

\begin{figure}[tb]
\epsfig{file= 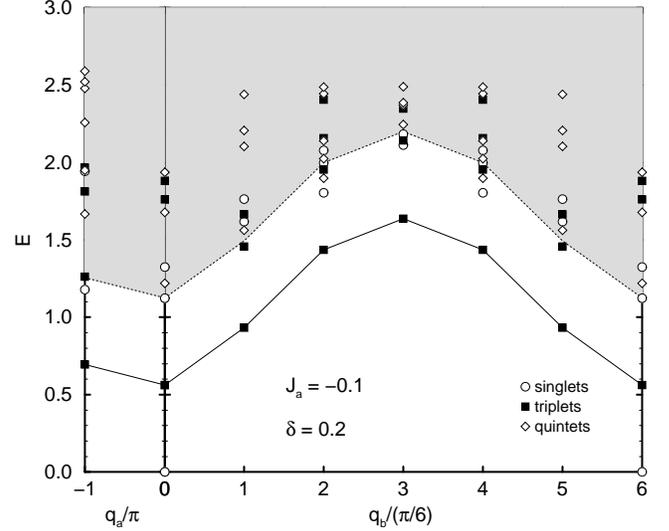, width = \linewidth, bbllx = 40, bblly = 40, bburx = 530, bbury = 450}
\caption{Low--energy excitations of two alternating chains coupled 
ferromagnetically; 
the parameters are: chain--length $=12, \delta = 0.2, J_a = -0.1$}
\label{figccdisp}
\end{figure}

Since an intra--chain frustration is not plausible in view on the structure 
of \vopo, we will consider a perpendicular coupling of two alternating chains 
instead:
\begin{eqnarray}\label{cc_ham}
  H_{\rm CC} & = & J_b \sum_{i\atop j = 1,2} 
  (1+\delta(-1)^i) {\bf S}_{i,j}\cdot{\bf S}_{i+1,j} \nonumber\\
  & & + J_a \sum_i {\bf S}_{i,1}\cdot{\bf S}_{i,2}.
\end{eqnarray}
Here $j$ is numbering the chains. As was already suggested 
in~\cite{Garrett2}, such a coupling in a--direction should be 
ferromagnetic to explain the observed dispersion. 
For illustration, in Fig.~\ref{figccdisp} we plotted a few low--lying 
energies of a $2\times 12$ system with $\delta = 0.2$ and $J_a = -0.1$.
Similar results were obtained for a $3\times 8$ system. 

Again we have a well--defined magnon branch. As a guide to the eye
we shaded the region where one would expect a two--magnon continuum, 
approximated here by adding the gap at zero momentum to the magnon 
dispersion. 
Close to the continuum edge there are several states: singlets, triplets, 
as well as quintets. To gain further insight one has to perform a
finite--size analysis. For extrapolation to the infinite system we use 
the following formulas for the lowest singlets and 
triplets~\cite{Bouzerar1,Barnes3}:
\begin{eqnarray}
  E^{\rm S}(L) & = & E^{\rm S}(\infty) + ({B\over L} + C) e^{-L/A}\label{sinscal}\\
  E^{\rm T}(L) & = & E^{\rm T}(\infty) + {B\over L} e^{-L/A}\label{triscal}
\end{eqnarray}

\begin{figure}[tb]
\epsfig{file= 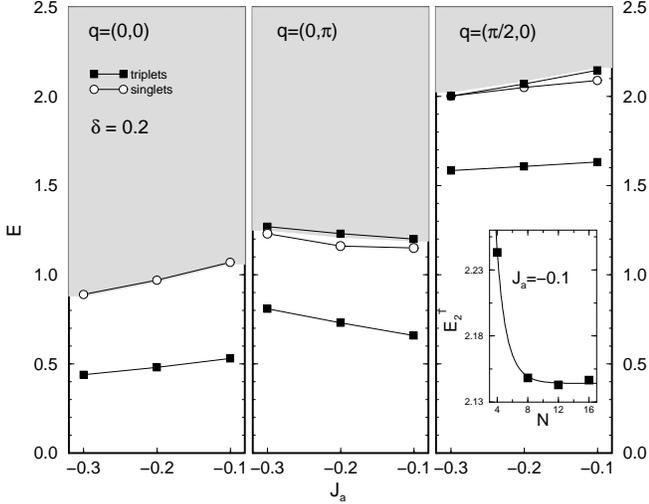, width = \linewidth, bbllx = 50, bblly = 50, bburx = 560, bbury = 450}
\caption{Energy of the lowest singlet (open circles) and triplet (filled squares)
excitations for different interchain coupling $J_a$ and fixed 
dimerization $\delta = 0.2$.}
\label{figccfit}
\end{figure}

In Fig.~\ref{figccfit} a few low--lying excitations at momenta 
$(q_b,q_a)=(0,0)$, $(0,\pi)$ and $(\pi/2,0)$ are given subject to the 
ferromagnetic interchain coupling $J_a$. At $(0,0)$ and
$(0,\pi)$ the data is extrapolated to infinite system size, while
at $(\pi/2,0)$ results for a $2 \times 16$ system are shown, because
we have just four different system sizes ($2 \times 4, 8, 12, 16$) at
this momentum, making finite--size scaling questionable.
Nevertheless, in the inset we tried to extrapolate the second triplet 
at $(\pi/2,0)$ to the infinite system, using the ansatz of 
Eq.~(\ref{triscal}). The plot indicates that this triplet shows a weak 
nonmonotonic behaviour, in contrast to the single chain case.

Obviously at momenta $(0,0)$ and $(0,\pi)$ there are no second
triplets below the two--magnon continuum. Just at $(q_b,q_a)=(\pi/2,0)$
a second triplet stays very close to the continuum edge, and the 
well--defined singlet excitation, known from the single alternating chain
seems to disappear with increasing interchain coupling $J_a$. 

From the above results we conclude that an interchain coupling of this
simple type does not qualitatively change the structure of the 
low--energy excitations compared to the single alternating chain. 
Excitations are just shifted (as it seems linearly with
$J_a$ in most cases), but no new features appear. This is, why
we propose another model for \vopo.

\begin{figure}[tb]
\epsfig{file= 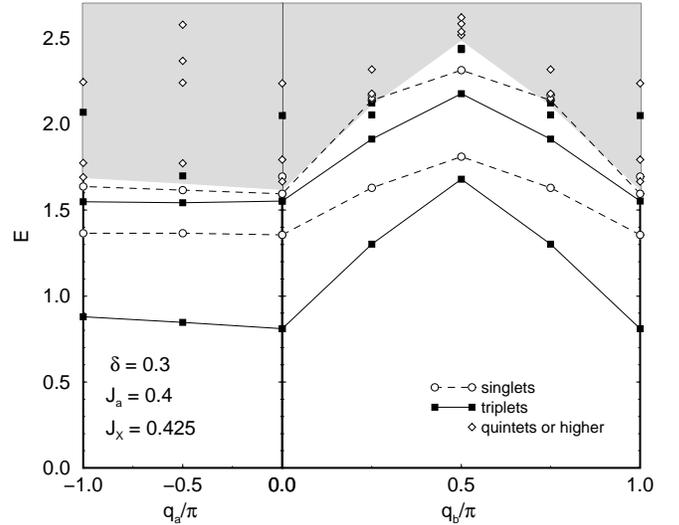, width = \linewidth, bbllx = 40, bblly = 40, bburx = 530, bbury = 450}
\caption{Low--energy excitations of the 2D model; the system size is
$4 \times 8$, $\delta = 0.3, J_a = 0.4$ and $J_{\times} = 0.425$}
\label{figXdisp}
\end{figure}

We mentioned above that frustration in the alternating chain
can lead to a well--defined triplet below the two--magnon
continuum. Thus going to the second dimension we include an additional,
frustrating coupling $J_{\times}$. Then our model Hamiltonian
reads
\begin{eqnarray}\label{cross_ham}
  \lefteqn{H_{\times} = J_b \sum_{i, j} (1+\delta(-1)^i) 
    {\bf S}_{i,j}\cdot{\bf S}_{i+1,j} }\nonumber\\ 
  & & + J_a \sum_{i,j} {\bf S}_{i,j}\cdot{\bf S}_{i,j+1}\\
  & & + J_{\times} \sum_{i,j} ({\bf S}_{2i,j}\cdot{\bf S}_{2i+1,j+1} +
  {\bf S}_{2i+1,j}\cdot{\bf S}_{2i,j+1})\nonumber
\end{eqnarray}
(cf. also Fig.~\ref{figstruct}, lower panel). 
As yet there is no data available about the strength of such a
coupling, but as a first step it seems not unreasonable in view of the
oxygen--mediated superexchange paths in \vopo. We assume all exchange 
integrals to be antiferromagnetic, but still the parameter space is very 
large.
It appears
that $J_\times$ has to be bigger than $J_a$ to get a ferromagnetic
magnon dispersion in a--direction, what is plausible. On the other hand,
both couplings should not differ too much for a second triplet
branch to exist in the whole momentum space, and should have a
sufficient strength. The size of the gap to the first triplet branch
is (still) mainly controlled by the dimerization $\delta$.

A good choice of parameters is $\delta = 0.3$, $J_a = 0.4$ and 
$J_\times = 0.425$, for which we diagonalized systems of two, three and 
four chains with a total number of up to 32 spins and periodic boundary 
conditions. The low--energy excitations of the $4 \times 8$ system are
shown in Fig.~\ref{figXdisp}.

Beside two triplet branches we observe also a well--defined
singlet, and there might even be a second singlet near momentum
$(q_b,q_a) = (\pi/2,0)$. As the difference between $J_\times$ and
$J_a$ is small, the dispersion of the triplets is weak in a--direction,
in accordance with experiments.
We stress that the picture remains
qualitatively unchanged going from the $3 \times 8$ to the
$4 \times 8$ system, just the second triplet shifts downwards at 
momentum $(0,0)$ with increasing system size. Thus we believe that 
these features will survive in the infinite system.

\begin{figure}[bt]
\epsfig{file= 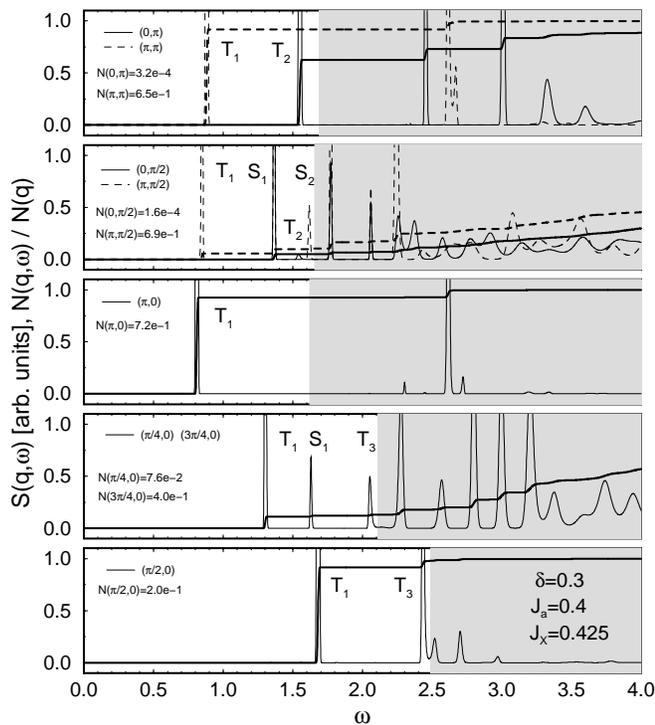, width = \linewidth, bbllx = 40, bblly = 70, bburx = 530, bbury = 630}
\caption{Dynamical spin structure factor for the 2D model ($4 \times 8$ system,
$\delta=0.3, J_a=0.4, J_{\times} = 0.425$); the lowest singlet and triplet
excitations are classified.}
\label{figXszq}
\end{figure}

To provide some more information on the excitation spectrum, we calculated the
dynamical spin structure factor and the integrated spectral weight
\begin{eqnarray}\label{struc_fac}
  S({\bf q},\omega) & = & \sum_n |\langle n|{\bf S}^z({\bf q})|0\rangle|^2 
  \delta(E_n - E_0 - \omega)\,,\\
  N({\bf q},\omega) & = & \int_0^{\omega} d\omega' S({\bf q},\omega')\,,
\end{eqnarray}
where ${\bf S}^z({\bf q}) = \sum_{i,j} e^{i {\bf q}\cdot{\bf r}_{i,j}} S^z_{i,j}$.
In the plot the integral is normalized to one; its real value 
$N({\bf q}) = N({\bf q},\infty)$ is noted in each panel.

It seems that we need a finite momentum component in b--direction 
for the first triplet to have some weight, while the second triplet
appears only in a--direction. For comparison take the dashed and solid lines
in the upper two panels of Fig.~\ref{figXszq}, corresponding to
momenta $(\pi,x)$ and $(0,x)$, respectively, that are equivalent in
energy.

To summarize, using exact diagonalization methods we have shown that a simple 
ferromagnetic coupling of alternating Heisenberg chains does not
provide two well--defined triplet branches as were observed in
inelastic neutron scattering experiments on vanadyl pyrophosphate \vopo. 
From our experience with frustrated alternating Heisenberg chains, we proposed
an alternative model to describe the low--energy physics of
\vopo, introducing a frustrating interchain coupling. Due to the
large parameter space and the computational effort for sufficiently
extended 2D systems, we made no attempt to fix the parameters for \vopo, 
but showed that the proposed model can describe the general feature of 
two triplet branches below a continuum of states. These triplets
exhibit a ferromagnetic (antiferromagnetic) dispersion in a-- (b--) direction.
Thus we believe that our model is a good starting point for further analysis.

We thank  E.~M\"uller-Hartmann, J.~Schliemann, S.~Sil and G.~Wellein for
stimulating discussion. G.B. acknowledges the hospitality at the Universit\"at
Bayreuth, granted by the Graduiertenkolleg 'Nichtlineare Spektroskopie und
Dynamik'.
Calculations were done on the T3E systems at HLRZ J\"ulich and HLRS 
Stuttgart and on the SP2 systems at GMD Bonn and LRZ M\"unchen.



\begin{references}

\bibitem{Johnston} D.C.~Johnston, J.W.~Johnson, D.P.~Goshorn, 
and A.J.~Jacobson, Phys. Rev. B {\bf 35}, 219 (1987).

\bibitem{Barnes} T. Barnes and J. Riera, Phys. Rev. B {\bf 50}, 6817 (1994).

\bibitem{Schwenki} A.V.~Prokofiev, F.~B\"ullesfeld, W.~Assmus, H.~Schwenk, 
D.~Wichert, U.~L\"ow, and B.~L\"uthi, cond-mat/9712137.

\bibitem{Eccleston} R.S. Eccleston, T. Barnes, J. Brody, and J.W. Johnson, 
Phys. Rev. Lett. {\bf 73}, 2626 (1994).

\bibitem{Garrett1} A.W. Garrett, S.E. Nagler, T. Barnes, and B.C. Sales, Phys. Rev.B {\bf 55}, 3631 (1997).

\bibitem{Garrett2} A.W. Garrett, S.E. Nagler, D.A. Tennant, B.C. Sales, and T. 
Barnes, Phys. Rev. Lett. {\bf 79}, 745 (1997). 

\bibitem{Bouzerar1} G.~Bouzerar, A.P.~Kampf, and G.I.~Japaridze, cond-mat/9801046 (accepted for publication in Phys. Rev. B); 
G. Bouzerar and S. Sil, cond-mat/9805042.

\bibitem{Barnes2} T.~Barnes, J.~Riera, and D.A.~Tennant, cond-mat/9801224.

\bibitem{Uhrig} G.S.~Uhrig and H.J.~Schulz, Phys. Rev. B {\bf 54}, 
R9624 (1996).

\bibitem{Barnes3} T.~Barnes, E.~Dagotto, J.~Riera, and E.S.~Swanson, 
Phys. Rev. B {\bf 47}, 3196 (1993).

\end{references}
\end{document}